\documentstyle[12pt]{article}
\begin{document}
\begin{flushright}
{\bf gr-qc/9911062}\\
{\bf UCVFC-DF/16-99}
\end{flushright}
\vskip 1truecm
\begin{center}
{\LARGE Analytical Description of Voids in Majumdar-Papapetrou Spacetimes}
\vskip .4truecm

{\bf Victor Varela}\footnote{Talk delivered at the Spanish Relativity
Meeting. Bilbao, Spain, September 7-10, 1999.}\\

{\it Department of Mathematics and Statistics,\\
The University of Edinburgh,\\
Edinburgh, Scotland\\
\vskip 12pt
Escuela de F\'{\i}sica, Facultad de Ciencias,\\
Universidad Central de Venezuela,\\
Caracas, Venezuela \footnote{Present address.}\\
{\em E-mail:{\tt varela@belka.ciens.ucv.ve}}\\
\vskip 12pt
Departamento de F\'{\i}sica,\\
Universidad Sim\'{o}n Bol\'{\i}var,\\
Caracas, Venezuela}
\end{center}

\abstract{
We discuss new Majumdar-Papapetrou solutions
for the 3+1 Einstein-Maxwell equations, with charged dust acting as
the external source of the fields. The solutions satisfy non-linear potential
equations which are related to well-known wave equations of 1+1 soliton
physics. Although the matter distributions are not localised, they present
central structures which may be identified with voids.}

\newpage

\section{Introduction}
We consider solutions for the Einstein-Maxwell (EM) equations with charged dust 
acting as the external source of the fields. Our basic equations read
\begin{equation}
G^{\mu}_{\nu}=8\pi T^{\mu}_{\nu},
\label{eq:eins}
\end{equation}
\begin{equation}
F^{\mu\nu}{}_{;\nu}=4\pi J^{\mu},
\label{eq:max}
\end{equation}
where $G^{\mu}_{\nu}$ and $F^{\mu\nu}$ denote
the Einstein and Maxwell tensors, and the total
energy-momentum tensor is given by
\begin{equation}
T^{\mu}_{\nu}=E^{\mu}_{\nu}+\rho u^{\mu}u_{\nu}.
\label{eq:temunu}
\end{equation}
Here $E^{\mu}_{\nu}$ is the Maxwell energy-momentum tensor, 
and the matter term corresponds to 
dust with energy density $\rho$ and four-velocity $u^{\mu}$.
The four-current is defined by the expression
\begin{equation}
J^{\mu}=\sigma u^{\mu},
\label{eq:current}
\end{equation}
were $\sigma$ is the charge density.

We assume that the fluid is static and use the conforstatic metric 
\begin{equation}
ds^{2}=-V^2\;dt^{2}+\frac{1}{V^2}\;h_{ij}dx^{i}dx^{j},
\label{eq:confor}
\end{equation}
where the background metric $h_{ij}$ and $V$ depend only on the space-like 
coordinates $x^{1}$, $x^{2}$, $x^{3}$.
The electrostatic forms of $A_{\mu}$ and $J^{\mu}$ are given by
\begin{equation}
A_{\mu}=A_{0}(x^{i})\delta^{0}_{\mu},
\label{eq:Astatic}
\end{equation}
\begin{equation}
J^{\mu}=\frac{\sigma (x^{i})}{V}\delta^{\mu}_{0},
\label{eq:Jstatic}
\end{equation}
with $i=1,2,3$.

Under these conditions, Eq. (\ref{eq:max}) contains only one non-trivial
equation:
\begin{equation}
\frac{1}{\sqrt{h}}\;\partial_{j}\left(\sqrt{h}\; h^{jk}\;
\frac{\partial_{k}A_{0}}
{V^{2}}\right)=\frac{4\pi J^{0}}{V^2},
\label{eq:nontriv}
\end{equation}
where $h$ and $h^{ij}$ are the determinant and the inverse of $h_{ij}$,
respectively.

The trace of the Einstein equations is
\begin{equation}
R=-8\pi T,
\label{eq:trace}
\end{equation}
where $R$ denotes the Ricci scalar and $T=T^{\mu}_{\mu}$. We use the 
decomposition
\begin{equation}
R=V^2\left[R_{h}+2 \nabla^{2}_{h}\ln{V}-2\partial_{i}\ln{V}\partial^{i}\ln{V}
\right].
\label{eq:Rdecom}
\end{equation}
Here $R_{h}$ is the Ricci scalar associated to $h_{ij}$,
and $\nabla^{2}_{h}$
is the three-dimensional Laplacian operator constructed with the same metric.
We assume a flat background space, with $R_{h}=0$. Therefore, combining 
Eqs. (\ref{eq:trace}) and (\ref{eq:Rdecom}) we obtain
\begin{equation}
\nabla^{2}_{h}\left(\frac{1}{V}\right)=\frac{4\pi T}{V^{3}}.
\label{eq:einslap}
\end{equation}
Following the Majumdar-Papapetrou (MP) procedure,\cite{ma,pa}
we assume that
\begin{equation}
A_{0}=\alpha V,
\label{eq:mapass}
\end{equation}
where $\alpha=\pm 1$. As a consequence, the Maxwell equation (\ref{eq:nontriv})
takes the form
\begin{equation}
\nabla^{2}_{h}\left(\frac{1}{V}\right)=-\frac{4\pi \alpha J^{0}}{V^{2}},
\label{eq:maxred}
\end{equation}
which is clearly the same as Eq. (\ref{eq:einslap}) whenever the condition
\begin{equation}
T=-\alpha J^{0} V
\label{eq:compa}
\end{equation}
holds. This equation can be combined with $J^{0}=\frac{\sigma}{V}$ to obtain
the alternative expression
\begin{equation}
\sigma=-\alpha T.
\label{eq:cden}
\end{equation}
Since $T=-\rho$ for dust, Eqs. (\ref{eq:einslap}) and (\ref{eq:cden})
can be finally expressed as
\begin{equation}
\nabla_{h}^{2}\lambda+4\pi \rho \lambda^{3}=0,
\label{eq:nonlinpot}
\end{equation}
\begin{equation}
\sigma=\alpha\rho,
\label{eq:sigmadust}
\end{equation}
where $\lambda=\frac{1}{V}$. 
Due to Eqs. (\ref{eq:confor}) and (\ref{eq:mapass}), only one Einstein equation 
is not trivially satisfied.\cite{gurses,ida}
Therefore, solving Eq. (\ref{eq:nonlinpot}) is sufficient for finding a solution
of the EM equations.

If we identify our flat background space with the Euclidean,
three dimensional space and assume $\rho=0$, then
Eq. (\ref{eq:nonlinpot})
reduces to the usual  Laplace equation $\nabla^{2}\lambda=0$ and the
electrovac, multi-black hole solution 
follows straightforwardly. Assuming spherical symmetry,
and using spherical coordinates, we find 
\begin{equation} 
\lambda=1+\frac{m}{r}.
\label{eq:esfesol}
\end{equation}
In the far-asymptotic region, the behaviour of this solution is approximately
given by
\begin{eqnarray}
V \approx 1-\frac{m}{r}, &
g_{00} \approx -1+\frac{2m}{r}, &
A_{0} \approx \pm (1-\frac{m}{r}).
\label{eq:assymp1}
\end{eqnarray}
The corresponding expression for the electric field is
\begin{equation} 
E \approx \frac{q}{r^2},
\label{eq:assymp2}
\end{equation}
where
\begin{equation} 
q=\pm m.
\label{eq:qigualm}
\end{equation}

Equation (\ref{eq:confor}) 
implies that
the invariant area of any 2-sphere surrounding the origin
is given by
$\frac{4\pi r^{2}}{V(r)^{2}}$. Therefore, 
the set $r=0$, $t=constant$
has a non-zero invariant area given by $4\pi m^{2}$.
In fact, a simple coordinate transform shows that the null hypersurface $r=0$
is the horizon of the extremal Reissner-Nordstr\"{o}m solution.
Also, if we define the new radial coordinate $\tilde{r}=-r$ 
and perform the standard analysis,\cite{haha}
then we find that this horizon encloses a point-like, essential singularity
placed at  $\tilde{r}=m$. In fact, the invariant area vanishes and the 
scalar
$J=F_{\mu\nu}F^{\mu\nu}=\lambda^{-4}\left(\frac{d\lambda}{dr}\right)^{2}$ 
blows up at that point.

Equations (\ref{eq:nonlinpot}) and (\ref{eq:sigmadust}) were originally 
discussed by Das \cite{das} in his study of equilibrium configurations of
self-gravitating, charged dust.
More recently, G\"{u}rses \cite{gurses} has considered
non-electrovac solutions when Eq. (\ref{eq:nonlinpot}) is linear.
This situation corresponds to his choice $\rho=\frac{b^{2}}{4\pi\lambda^{2}}$
for constant $b$.
In this case, Eq. (\ref{eq:nonlinpot}) admits the particular solution
$\lambda=\frac{a\sin br}{r}$
where $a$ is an integration constant. The oscillatory behaviour of this solution
implies a geometry with a complicated radial dependence. In fact, the
invariant area vanishes for a discrete, infinite set of values of $r$, and the 
Ricci scalar $R=\frac{2 b^2 r^2}{a^2 \sin^{2}br}$ blows up wherever the
invariant area vanishes, except for $r=0$. Other solutions with oscillatory 
behaviour have been considered by Balakrishna and Wali,\cite{bawa} 
Braden and Varela,\cite{brava} and Ida.\cite{ida}
In Section 2 we exploit
the general non-linearity of
Eq. (\ref{eq:nonlinpot}) to obtain new solutions which are free of
oscillatory singularities and allow asymptotically flat behaviour.

\section{The non-linear models}
The non-linear potential equation (\ref{eq:nonlinpot}) takes the spherically
symmetric form
\begin{equation}
\frac{d^{2}\lambda}{dr^{2}}+\frac{2}{r}\frac{d\lambda}{dr}+4\pi\rho\lambda^{3}=0.
\label{equesfe}
\end{equation}
Using the new radial coordinate $\tau=\frac{1}{r}$, the same differential
equation can be written as
\begin{equation}
\frac{d^{2}\lambda}{d\tau^{2}}+\frac{4\pi\rho}{\tau^{4}}\lambda^{3}=0.
\label{equesfetau}
\end{equation}
If $\rho$ and $\lambda$ satisfy the condition
\begin{equation}
\rho=\frac{b^2}{4\pi}\frac{\tau^{4}\sin\lambda}{\lambda^{3}},
\label{newfun}
\end{equation}
then (\ref{equesfetau}) finally reduces to the
-sine-Gordon equation \cite{saha}
\begin{equation}
\frac{d^{2}\lambda}{d\tau^{2}}+b^2 \sin\lambda=0,
\label{soliton0}
\end{equation}
which has the solutions
\begin{equation}
\lambda^{\pm}\left(\tau\right) = 
2\arcsin\left[\tanh\left(\pm b\tau+c\right)\right]+2n \pi,
\label{newsol1}
\end{equation}
where $n$ is an arbitrary integer, $c$ is an integration constant,
and $b$ is assumed to be positive. We consider only the case $n=0$.
In terms of the original radial coordinate, these solutions read
\begin{equation}
V^{\pm}\left(r\right)=
\frac{1}{2\arcsin\left[\tanh\left(\pm \frac{b}{r}+c\right)\right]}.
\label{newsol2}
\end{equation}

We observe that $V^{\pm}(0)^{2}$ is finite, so the
invariant area vanishes for $r=0$. Therefore, the set $r=0$, $t=constant$ is
point-like with respect to both solutions. Let us deal with $V^{+}$ first.
A preliminary numerical study of the invariants $J$, $R$,
$R^{\alpha\beta}R_{\alpha\beta}$, 
$R^{\alpha\beta\gamma\delta}R_{\alpha\beta\gamma\delta}$
suggests that these 
quantities are bounded for non-negative $r$, whenever $c$ is positive. 
If we choose
\begin{equation}
c=\frac{1}{2}\ln\left[\frac{1+\sin(1/2)}{1-\sin(1/2)}\right],
\label{choice}
\end{equation}
then the far-asymptotic behaviour of this solution is given by 
Eqs. (\ref{eq:assymp1}), (\ref{eq:assymp2}), (\ref{eq:qigualm})
with $m=2b\cos(1/2)$. Therefore, $V^{+}$
is asymptotically flat, exactly as the MP electrovac solution.

The positive definite energy density given by Eq. (\ref{newfun}) corresponds
to a non-localised matter (and charge) distribution. However, $\rho$ is
negligible for $x=\frac{r}{b} \ll 0.2$. 
For very small $x$ the dimensionless expressions of $\rho$ and 
$R^{\alpha\beta\gamma\delta}R_{\alpha\beta\gamma\delta}$ are 
approximately given by
\begin{equation}
\rho(x)\approx \frac{e^{-c}}{\pi^{4}}\;\frac{e^{-\frac{1}{x}}}{x^{4}},
\label{approxrho}
\end{equation}
\begin{equation}
R^{\alpha\beta\gamma\delta}R_{\alpha\beta\gamma\delta}(x)
\approx 3\left(\frac{2}{\pi}\right)^{6}e^{-2c}\;\frac{e^{-\frac{2}{x}}}{x^{8}}.
\label{approxRR}
\end{equation}
These results imply a very fast decrease of $\rho (x)$ and 
$R^{\alpha\beta\gamma\delta}R_{\alpha\beta\gamma\delta}(x)$
when $x\rightarrow 0^{+}$, and suggest the existence of a void in the 
innermost region of the asymptotically flat object constructed with $V^{+}$.
Nevertheless, this interpretation cannot be complete without a better 
understanding of the point-like set $r=0$, $t=constant$.
A closer look at the
singularity contained in this solution is also necessary.
We interpret $r=0$, $t=constant$ as the center of symmetry and observe that
the above mentioned invariants are bounded at this point.
However, the coordinate transform $\tilde{r}=-r$ reveals the existence of a
point-like, essential singularity at $\tilde{r}=\frac{b}{c}$. In fact, the
invariant $J$ blows up at this point. The use of $V^{+}$ alone may imply the
division of the manifold into connected parts, separated by the point-like
singularity placed at $r=-\frac{b}{c}$. However, 
a very different situation comes out
when we restrict $V^{+}$ to positive values of $r$ and describe 
the geometry for $r<0$
with the second solution $V^{-}$. Then, a smooth (at least $C^{1}$)
matching of $V^{+}$ and $V^{-}$ occurs at $r=0$ and the arising 
asymptotically flat
spacetime seems to be connected and singularity free, 
with an almost empty region near
the center of symmetry. Thus, the joint use of $V^{+}$ 
and $V^{-}$ provides a simpler
description of a MP void. 
The study of the global structure of these solutions is left as
an open problem, which provides motivation for further research work.

Finally, we point out that
other exact, non-linear solutions for this theory can be found if we impose
different relationships between $\rho$ and $\lambda$. For example, the choice
\begin{equation}
\rho=-\frac{b^2}{4\pi}\frac{\tau^{4}\sin\lambda}{\lambda^{3}}
\label{newfun1}
\end{equation}
leads to the sine-Gordon equation
\begin{equation}
\frac{d^{2}\lambda}{d\tau^{2}}=b^2 \sin\lambda.
\label{sg}
\end{equation}
It has the well-known solutions
\begin{equation}
\lambda^{\pm}(\tau)=4 \arctan e^{(\pm b\tau+d)}.
\label{sgsol}
\end{equation}
If we choose $d=\ln \left[\tan(1/4)\right]$, 
then both solutions have asymptotically
flat behaviour.

Another example is
\begin{equation}
\rho=\frac{b^{2}}{4\pi}\left(\lambda-\lambda^{3}\right).
\label{newfun2}
\end{equation}
In this case the geometry is determined by the $\lambda\phi^{4}$ equation
\begin{equation}
\frac{d^{2}\lambda}{d\tau^{2}}+b^2\left(\lambda-\lambda^{3}\right)=0
\label{lphi4}
\end{equation}
which admits the solutions 
\begin{equation}
\lambda^{\pm}(\tau)=\tanh\left(\pm \frac{b}{\sqrt{2}}\tau+f\right).
\label{lphi4sol}
\end{equation}

The relationship between the 3+1 EM theory and the equations of 
1+1 soliton physics
deserves a more detailed examination. Possible extensions of this
work involve the analysis of dust models for which $\lambda(\tau)$ 
is a solution of the KdV
equation, and the study of the non-linear potential equations arising in
higher dimensions.\cite{mitra}

\vspace*{-2pt}

\section*{Acknowledgments}
The author is grateful to the Mathematics and Statistics Department      
of Edinburgh University for its hospitality
during his long-term
research visit. 
He is indebted to Professors Nikolaos Batakis (Ioannina)
and Luis Herrera (Salamanca) 
for useful discussions and comments.
Finally, he thanks Dr. Pio J. Arias (UCV) and Professor Jorge Zanelli 
(San\-tiago) for valuable suggestions,
and the {\it Grupo de Relatividad y Campos} at Sim\'{o}n 
Bol\'{\i}var University for partial financial support.

\vspace*{-9pt}


\end{document}